\documentclass[aps,prb,twocolumn,showpacs,preprintnumbers]{revtex4}
\usepackage{amssymb}
\usepackage{multirow}
\usepackage[colorlinks,linkcolor=blue,citecolor=blue,dvipdfm]{hyperref}
\usepackage{graphicx}
\usepackage{dcolumn}
\usepackage{bm}
\usepackage{color,soul}

\begin{document}

\title{The commensurate magnetic excitations induced by band-splitting and Fermi surface topology in n-type Cuprates}

\begin{abstract}
The antiferromagnetic correlation plays an important role in high-T$_{c}$ superconductors. Considering this effect, the magnetic excitations in n-type cuprates near the optimal doping are studied within the spin density wave description. The magnetic excitations are commensurate in the low energy regime and further develop into spin wave-like dispersion at higher energy, well consistent with the inelastic neutron scattering measurements. We clearly demonstrate that the commensurability originates from the band splitting and Fermi surface topology. The commensurability is a normal state property, and has nothing to do with d-wave superconductivity. The distinct behaviors of magnetic excitation between the n-type and p-type cuprates are further discussed. Our results strongly suggest the essential role of antiferromagnetic correlations in the cuprates.

\end{abstract}

\author{H. Y. Zhang$^1$, Y. Zhou$^1$, C. D. Gong$^2,^1$, and H. Q. Lin$^3$ }
\affiliation{$^1$National Laboratory of Solid State Microstructure, Department of
Physics, Nanjing University, Nanjing 210093, China \\
$^2$Center for Statistical and Theoretical Condensed Matter Physics,
Zhejiang Normal University, Jinhua 321004, China\\
$^3$Department of Physics and the Institute of Theoretical Physics, Chinese
University of Hong Kong, Hong Kong, China}
\maketitle
\date{\today }

\section{Introduction}

The parent compounds of the high-$T_{c}$-temperature superconductors
are antiferromagnetic (AFM) Mott insulator. Superconductivity (SC) emerges
when charge carries (holes or electrons) are doped into the $CuO_{2}$
planes. As well known that the clear electron-hole asymmetry is found in the
phase diagram. For hole doped case, the AFM and superconducting phases are separated by spin glass phase. In contrast, the AFM phase extends
over a much wider range of doping and even coexists with SC in the electron doped cuprates\cite{Armitage-RMP10}. Due to proximity
of antiferrmagnetism and SC, it is generally believed that there exists
intrinsic link between these two phases. The studies of the spin dynamics in n-type and p-type cuprates will shed light on the mechanism of superconductivity.

One of the most available techniques to study the spin dynamics is the inelastic neutron scattering
(INS), which directly measures the magnetic excitations (MEs). Compared with the well studied p-type cuprate\cite{Hourglass,p-type-THEO}, the investigations on the MEs in n-type cuprates\cite{Yamada-PRL03,Wilson-Nature06,Motoyama-Nature07,Fujita-PRL08,Fujita-JPCS08,Zhao-NPHY11} are much less due to the technical reason. A robust feature of MEs in n-type
cuprates, i.e., the commensurate spin response, had been revealed by these INS measurements. The commensurability, characterized by the strongest intensity peaked at $\textbf{Q}=(\pi,\pi)$, covers for a wide low-energy region near the optimal doped NCCO\cite%
{Yamada-PRL03} and PLCCO\cite{Wilson-Nature06}. Further detecting shows that
such commensurability in n-type cuprates exists for a wide doping range from
the underdoping to heavy overdoping\cite{Fujita-PRL08,Fujita-JPCS08}. More importantly, the commensurate MEs persist well above superconducting critical temperature $T_{c}$, indicating its non-superconducting origin. It
gradually develops into the spin-wave-like dispersion centered around the $Q$
point at higher energy, analogous to its undoped parent compound\cite%
{Wilson-Nature06}. In contrast, The well known 'hourglass' type magnetic
dispersion had been discovered in the p-type cuprates, where the commensurate peak
can only be found at the resonance energy. Therefore, the two types of cuprates exhibit distinct spin
response, indicating the intrinsic particle-hole asymmetry.

Theoretically, several works had been carried out to interpret the MEs
features of n-type cuprates. Adopting the single band description with experimentally fitted parameters, Kr\"{u}ger \textit{et al.} claimed that the fermiology random
phase approximation (RPA) approach with a momentum independent (or weakly
dependent) Coulomb repulsion cannot account the low-energy commensurability. Their numerical results indicated that the MEs in n-type
case should be more incommensurate than that in p-type case\cite%
{Kruger-PRB07}. Such conclusion may reveal the fact that the single band description is invalid in the n-type cuprates\cite{Fujita-PRL08}. Ismer
\textit{et al.} showed that this may be improved by a strongly momentum
dependent Coulomb repulsion with form of $U_{q}=U_{0}(\cos q_{x}+\cos q_{y})$%
\cite{Ismer-PRL07}. However, such improvement is more likely originated from
its sharply peaked form at $(\pi,\pi)$, which cannot be understood
physically. Using a slave-boson mean-field approach, Li \textit{et al.}
showed that the commensurability can be established in the SC state\cite%
{LJX-PRB03}. These above mentioned theoretical works are all based on the
belief that the d-wave SC is response for the low energy commensurability.
It is hard to image that the small superconducting gap can produce the wide energy range commensurability.
The most important is that the commensurate phenomenon is also found in the normal state of
n-type cuprates, where the d-wave SC disappears. The MEs had also been discussed within the frame work of coexisting of SC and AFM\cite{YQS-PRB05}. Commensurate ME had been subsequently obtained\cite{LJX-JPC10}. However, this result clearly contradicts with the stoner critition\cite{ROWE-PRB12,Schrieffer-PRB88}.
Furthermore, recent INS data reveal a magnetic quantum critical point where the
SC first appears, implying that the coexistence may not exist\cite%
{Motoyama-Nature07}. Additionally, as we mentioned above, the
commensurability also exists in the state without long-range AFM order.

In this paper, we focus on the commensurate MEs in the
n-type cuprates near the optimal doping. Its energy region is closely related the to strength of
effective $(\pi,\pi)$-scattering. The commensurate peak disappears at magnetic resonance energy $\omega^{M}_{res}$ and
develops into the spin-wave like dispersion. These features are qualitatively consistent with INS measurements. This commensurability is a normal state property, and has nothing to do with the superconductivity. We
explicitly demonstrate that the commensurability is originated from the
band splitting and Fermi surface topology. Therefore, the AFM correlation plays key roles in the n-type cuprates. The differences of MEs between the p-type and n-type cuprates are further discussed for comparison.


A spin-density wave (SDW) description is adopted to investigate the n-type cuprates near the
optimal doping. Such description is first suggested by Armitage \textit{et
al.} based on the ARPES measurements on $NCCO$\cite{Armitage-PRL01}. The underlying Fermi surface
disappears around the hot-spot near the optimal doping, where the long-range
antiferromagnetism is absent, strongly suggesting the existence of a $%
\mathbf{{Q}=(\pi, \pi)}$-scattering. Parker \textit{et
al.} further proposed an effective energy band with $\xi
_{k}^{\eta}=\epsilon _{k}^{\prime } +\eta \sqrt{\epsilon_{k}^{2}+V_{\pi ,\pi
}^{2}}$ ($\eta=1$, and $-1$ for upper, and lower band, respectively)\cite%
{Parker-PRB07}, where $V_{\pi,\pi}$ is the strength of the effective $\mathbf%
{Q}$-scattering, representing the influence of the SDW. $\epsilon_{k}$ and $%
\epsilon_{k}^{\prime}$ is the inter- and intra-lattice hopping term. This
description well reproduces the $\sqrt{2} \times \sqrt{2}$ band folding and
Fermi surface reconstruction\cite{Ikeda-PRB09,Matsui-PRL05}, and its
applications on the temperature evolution of optical conductivity\cite%
{Zimmer} and the Hall coefficient\cite{Dagon-PR} give qualitative agreement with experiments.
Now, the model Hamiltonian is expressed as
\begin{eqnarray}
H&=&\sum_{k\sigma }\epsilon _{k}^{\prime }(d_{k\sigma }^{+}d_{k\sigma
}+e_{k\sigma }^{+}e_{k\sigma }) +\sum_{k\sigma }\epsilon _{k}( d_{k\sigma
}^{+}e_{k\sigma }+hc.)  \nonumber \\
&&-\sum_{k\sigma }\sigma V_{\pi ,\pi }( d_{k\sigma }^{+}d_{k\sigma
}-e_{k\sigma }^{+}e_{k\sigma })\text{,}
\end{eqnarray}
where, the two sublattices $D$ and $E$ with respective fermionic operator $d$
and $e$ are introduced due to SDW\cite%
{EXPLAIN}. $\epsilon _{k}=-2t(cosk_{x}+cosk_{y})$, and $\epsilon
_{k}^{\prime}=-4t^{\prime}cosk_{x}cosk_{y}-2t^{\prime%
\prime}(cos2k_{x}+cos2k_{y})-\mu$ with $t$, $t^{\prime}$, and $%
t^{\prime\prime}$ are the fitting parameters for nearest-neighbor (NN),
second-NN, and third-NN hoping. The summation is restricted in the AFM
Brillouin zone. The quasiparticle dispersion $\xi_{k}^{\eta}$ can be obtained by the
rotation transformation, with corresponding weight factor $W^{\eta}=\frac{1}{2}(1+\eta
sin2\theta_{k})$. Here, $cos2\theta_{k}=\frac{V_{\pi ,\pi }}{\sqrt{\epsilon_{k}^{2}+V_{\pi ,\pi
}^{2}}}$, and
$sin2\theta_{k}=\frac{\epsilon
_{k}}{\sqrt{\epsilon_{k}^{2}+V_{\pi ,\pi }^{2}}}$.

Here, we would like to emphasize that the long-range AFM
order disappears near the optimal doping. As pointed by Motoyama \textit{et
al.} that the Neel temperature detected above $x=0.134$ in $NCCO$ originates
from the region of samples that were not fully oxygen-annealed\cite%
{Motoyama-Nature07}. This means that the genuine long-range
antiferromagnetism does not coexist with superconductivity. However, the $2$-dimensional AFM correlation remains. Unlike only several
lattice-distant length in p-type cuprates\cite{Kastner-RMP98}, the AFM
correlation is about tens lattice-distance in the optimal electron-doped
cuprates\cite{Motoyama-Nature07}. In this sense, the AFM
correlations in the n-type cuprates is similar to the long-range AFM order at
least in the small scaling. Therefore, using a slowly fluctuating SDW order
to describe the long-range AFM correlation is a considerable treatment.
Though present SDW description is analogous to the form in the AFM phase\cite%
{YQS-PRB05}, the physics behind is essentially different.

The spin susceptibility under random phase approximation is
\begin{equation}
\chi _{q}\left( \omega \right) =\frac{\chi _{q}^{0}- U\left( \chi
_{q}^{0}\chi _{q+Q}^{0}-\chi _{q,q+Q}^{0}\chi _{q+Q,q}^{0}\right) }{\left(
1-U\chi _{q}^{0}\right) \left( 1- U\chi _{q+Q}^{0}\right) -U^{2}\chi
_{q,q+Q}^{0}\chi _{q+Q,q}^{0}}\text{,}  \label{E2}
\end{equation}
with $U$ is a reduced Coulomb interaction due to the screening effect\cite%
{Zhou-PLA10}. The bare spin susceptibilities are
\begin{eqnarray}
\chi _{q,q}^{0} &=&\sum_{k}\sin ^{2}\left( \theta _{k+q}+\theta _{k}\right) \left( F_{--}+F_{++}\right) \nonumber \\
&&+\sum_{k}\cos ^{2}\left( \theta _{k+q}+\theta _{k}\right) \left(
F_{-+}+F_{+-}\right) \nonumber \\
\chi _{q,q+Q}^{0} &=&\sum_{k}(\cos2\theta_{k}-\cos2\theta_{kq})(F_{--}-F_{++}) \nonumber \\
&&-\sum_{k}(\cos2\theta_{k}+\cos2\theta_{kq}) \left( F_{-+}-F_{+-}\right)
\end{eqnarray}
with $F_{\eta \eta ^{\prime }}$ is
\begin{eqnarray}
F_{\eta \eta ^{\prime }} &=&\frac{1}{4}\left( f_{kq}^{\eta }-f_{k}^{\eta ^{\prime
}}\right) \left( \frac{1}{\omega -\xi_{kq}^{\eta }+\xi_{k}^{\eta ^{\prime }}}\right)\text{,}
\end{eqnarray}
where $f_{k}=1/(1+e^{\xi_{k}/kT)}$ is the Fermi distribution function. In numerically, the doping level is fixed at $x=0.15$, near the
AFM quantum critical point\cite{Motoyama-Nature07}. $t=250%
\emph{meV}$, $t^{\prime }=-50\emph{meV}$, and $t^{\prime \prime }=20\emph{meV}$ are adopted%
\cite{Parker-PRB07}. The best fitted effective $\mathbf{{Q}-}$ scattering
strength is $V_{\pi ,\pi }=100\emph{meV}$, and it will be adjusted for necessary. The
temperature is fixed at $T=0.2\emph{meV}$. We adopt a broaden factor $%
\Gamma $ to calculate the spin susceptibility. The reduced Coulomb
interaction is about $600\emph{meV}\sim 760\emph{meV}$, which is about $2\sim3t$. Our calculations are carried out on a mesh with $2048 \times 2048$ k-point in the full Brillouin zone.

\begin{figure}[tbp]
\vspace{0.00in} \hspace{-0.3in} \centering%
\includegraphics[width=3.5in]{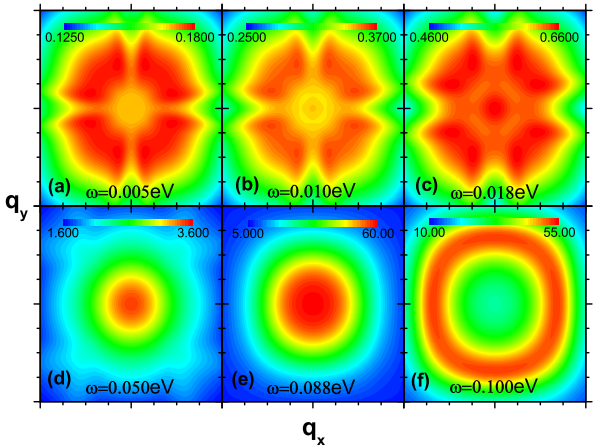} \vspace{-0.3in}
\caption{The typical momentum distribution of MEs $\Im
\protect\chi _{q}(\protect\omega)$ for different energy $\protect\omega$.
Only the near $\mathbf{Q}$ region are shown for clarification with $\frac{15%
\protect\pi}{16}\le q_{x} (q_{y}) \le \frac{17\protect\pi}{16}$. $U=700\emph{meV}$,
the effective $\mathbf{(Q)}$-scattering strength $V_{\protect\pi,\protect\pi%
}=100\emph{meV}$, and the broaden factor $\Gamma=10\emph{meV}$.}
\label{f.1}
\end{figure}

The typical energy-evolution of the MEs $\Im \chi_{q}(\omega)$ is shown in Fig.~\ref{f.1}. In the low-energy regime below $18\emph{meV}$ (Fig.~\ref{f.1}(a),
(b)), the MEs are incommensurate with strong intensity at
diagonal directions. Simultaneously, the intensity near $\mathbf{Q}$
enhances gradually. In the intermediate-energy regime, the strongest
intensity locates at the $\mathbf{Q}$ point, leading to the the so-called
commensurability (Fig.~\ref{f.1}(c), (d)). It maintains up to an critical
energy about $88\emph{meV}$ (Fig.\ref{f.1}(e)), where the strongest intensity $\Im
\chi_{\mathbf{Q}}(\omega)$ in the normal state can be found, and is referred as
the magnetic resonance $\omega_{res}^{M}$. The total energy range for the
commensurability is approximately $70\emph{meV}$ for $\Gamma=10meV$. This magnetic resonance is
directly related to the fact that the real part in
denominator of the RPA formula (Eq.~\ref{E2}) reduces to zero. Subsequently,
it evolves into a ring-like incommensurability in the high-energy region
with its radius expanding upon the further increased energy (Fig.~\ref{f.1}%
(f)). For high enough energy, the MEs are incommensurate with
strong intensity at the vertical directions (not shown).

\begin{figure}[tbp]
\vspace{0.00in} \hspace{-0.3in} \vspace{-0.0in} \centering%
\includegraphics[width=3.6in]{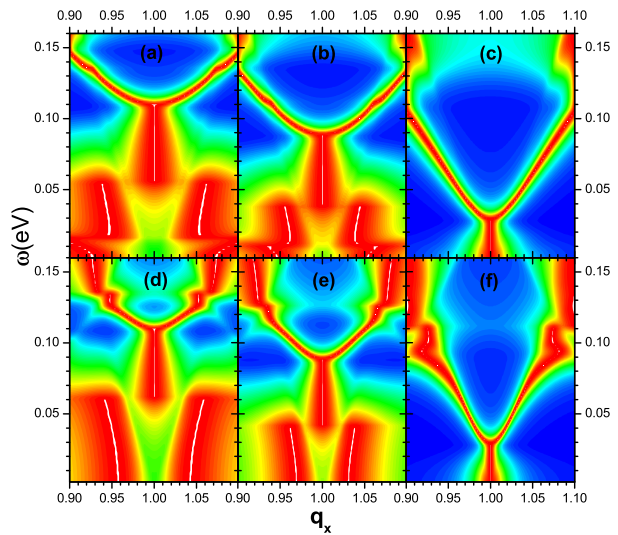} \vspace{-0.3in}
\caption{The dispersion of the MEs $\Im \chi_{q}(\omega)$ along the high symmetry direction. Upper panels are for
vertical direction with $q_{y}=\protect\pi$, and lower panels are for the
diagonal direction with $q_{y}=q_{x}$. From left to right is $U=660\emph{meV}$, $%
700\emph{meV}$, and $760\emph{meV}$, respectively. The effective $\mathbf{Q}$-scattering
potential is $V_{\protect\pi,\protect\pi}=100\emph{meV}$ and the damping rate $%
\Gamma=5\emph{meV}$. All data had been renormalized by setting the strongest
intensity at given $\omega$ as unit, denoted by the white lines.}
\label{f.2}
\end{figure}

Such features can be more clear in dispersion of the MEs at
high symmetry scanning lines as shown in Fig.~\ref{f.2}. A wide energy
regime with commensurability exists for all selected $U$, manifesting its
universal nature. The low-energy incommensurability increases slightly upon $%
\omega$. Hence the commensurability cannot not be viewed as the overlap of
two incommensurate peaks. It is an intrinsic feature of n-type cuprates. The low-energy incommensurability may be suppressed and even
absent with enhanced $U$. For example, when $U=0.76V$ (Fig.~\ref{f.2}(c)),
the MEs are still commensurate at low enough energy.
Correspondingly, the magnetic resonance energy $\omega_{res}^{M}$ decreases
down to $30\emph{meV}$ for $\Gamma=5meV$. The experimental discovered commensurability in $NCCO$\cite%
{Yamada-PRL03} and $PLCCO$\cite{Fujita-PRL08,Fujita-JPCS08,Fujita-JPSJ06}
near the optimal doping is more like similar to this case. The value of $%
U=760\emph{meV}$ is near the AFM stability, consisting with
the fact that the optimal doping is near the AFM quantum
critical point\cite{Motoyama-Nature07}.

The possible energy range of commensurability is mainly determined by the
effective $\mathbf{Q}$-scattering potential $V_{\pi,\pi}$. For $V=100\emph{meV}$ and
$\Gamma=5\emph{meV}$, it is about $48\emph{meV}$. This energy range decreases down to $10\emph{meV}$ when the $V=50\emph{meV}$. However, the realistic energy range of
commensurability may be substantially reduced for strong $U$ due to the proximity of the
AFM stability. It is only $30\emph{meV}$ for stronger $U=760\emph{meV}$.
For strong enough $U \ge 770\emph{meV}$ at given $V=100\emph{meV}$, the commensurability is
entirely suppressed and only the ring-like magnetic feature remains. This
situation is indeed an AFM state. Therefore, the ring-like
feature at high-energy regime in the electron-doped cuprates shares the same
origin as that in their parents compounds. In fact, those theories based on
the long-range AFM order\cite{LJX-JPC10,YQS-PRB05} cannot
account the commensurability found in $NCCO$\cite{Yamada-PRL03} and $PLCCO$%
\cite{Fujita-PRL08} due to Stone instability at $\omega=0$\cite{Schrieffer-PRB88}, unless
some special control parameter is adopted. The energy range of
commensurability also depends on the broaden factor $\Gamma$ as comparing
the data in Fig.~\ref{f.1} and Fig.~\ref{f.2} ((e) and (f)). However, this
phenomenon is still present even a small $\Gamma=1\emph{meV}$ is adopted,
which is less than the instrument resolution. Hence, the
commensurability is an intrinsic and universal property of the
electron-doped cuprates in the normal state.

The main difference in present work from the previous theoretical
investigation by Kr\"{u}ger \textit{et al.}\cite{Kruger-PRB07} is the influence of the AFM correlation is taken into account. The SDW description take the place of the single band description, producing a splitting
two-band. Therefore, the commensurability is a directly result of the
band-splitting. This can be also seem from the fact that commensurate energy
region diminishes with the reduced $V_{\pi,\pi}$ as we shown before. As we known that the AFM correlation weakens with doping\cite{Zhou-PRB08}. In the heavy overdoping range, the AFM correlation disappears, i.e., $V_{\pi,\pi}=0$, leading to the absence of band-splitting. Our results is then same as the work of Kr\"{u}ger \emph{et al.}, the MEs become incommensurate, consisting with INS measurements\cite{Fujita-PRL08}.

\begin{figure}[tbp]
\vspace{0.00in} \hspace{-0.3in} \vspace{-0.0in} \centering%
\includegraphics[width=3.6in]{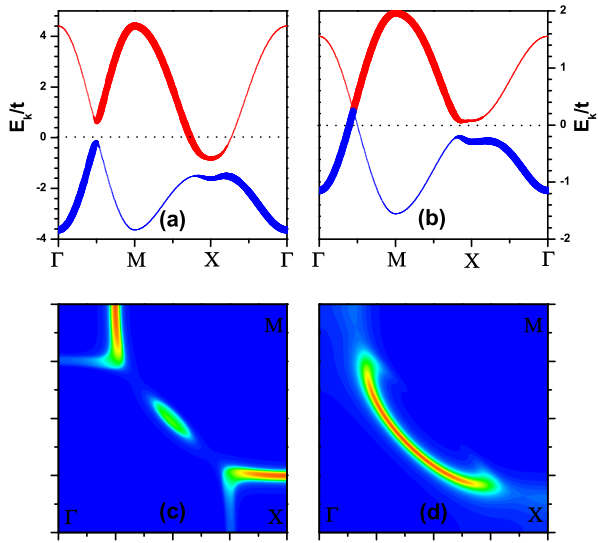} \vspace{-0.3in}
\caption{The electron sturecture (upper panels) and Fermi surface (lower panels). (a), and (c) are for n-type cuprates with SDW description described in the text. (b), and (d) are for p-type cuprates with YRZ ansatz\cite{Carbotte-PRB10}. The line thickness in (a), and (b) denotes the weight factor. The doping density is fixed at $x=0.15$.}
\label{f.3}
\end{figure}

In the p-type cuprate, the MEs exhibit the well known 'hourglass' dispersion, the commensurate peak emerges at the single energy point $\omega_{res}$. The underdoped p-type cuprates can also be described by the two-band description due to the opening of pseudogap\cite{Hole-twoband}, where the two bands near the antinodes are separated. Why do the two types of cuprates show significant particle-hole asymmetry? As well known that the Fermi surface is an 'arc' or hole pocket near the node point in the
underdoped p-type cuprates\cite{arc}, while it is an electron-pocket near antinodal point in the n-type
cuprates even near the optimal doping as shown in Fig.~\ref{f.3}. In fact, the two bands coincides at node point
in p-type cuprates, which leads to the single-point commensurability at resonance energy.
Furthermore, this can also interpret the commensurability exists in slightly overdoped n-type
cuprates even the large-three-pieced Fermi surface\cite{Armitage-PRL01} forms because of the band splitting at nodes.
Therefore, both the band splitting and Fermi surface topology play the important roles in the universal
commensurability in the n-type cuprates. As we stressed before, they both originate from the AFM correlation. Together with the previous theoretical works on the band
structure\cite{Ikeda-PRB09,Matsui-PRL05} and transport properties\cite{Zimmer,Dagon-PR},
we conclude that the AFM correlation plays essential roles in the cuprates.

The commensurate MEs remain in the presence of superconductivity. We introduce a phenomenological BCS-like pairing
term $-\sum_{k}\Delta _{k}\left( d_{k\uparrow }e_{-k\downarrow
}+e_{k\uparrow}d_{-k\downarrow }+h.c.\right) $ with standard $d$-wave symmetry
$\Delta_{k}=\Delta(\cos k_{x}-\cos k_{y})$. The resultant MEs change little, consistent with the INS observations. Therefore, the commensurability in n-type cuprate is a normal state property, and has nothing to do with SC.
The commensurability had been also obtained in a single band description with d-wave
superconductivity\cite{Ismer-PRL07,LJX-PRB03}. Though the d-wave
pairing produces two bands. However, the superconducting
gap in the optimal doped n-type cuprates is only $3\sim4 meV$\cite{Shan-PRB08,Dagan-PRL05}, too small to
account for the wide energy range commensurability. It seems that the commensurability comes from the
strong peaked factor $U_{q}$\cite{Ismer-PRL07} or $J_{q}$\cite{LJX-PRB03}
rather than the $d$-wave superconductivity in these theoretical
investigations. More importantly, the commensurability is a normal state
property, which can also be discovered well above the superconducting
transition temperature $T_{c}$.

In conclusion, the magnetic excitations near the optimal doped n-type
cuprates are studied within a spin-density wave description. The main features of magnetic excitations in the normal state are well
established. Our analyses clearly demonstrate that the band
splitting and the Fermi surface topology are the key for commensurability in n-type cuprates. This strongly suggests that the antiferromagnetic correlation plays important roles in cuprates. We emphasize that the
commensurability is a normal state property, and has nothing to do with
superconductivity. The qualitative agreement between the theoretical
calculations and experimental data also suggests the validity of the spin-density wave
description near the optimal doping where the long range
antiferromagnetic order is absent. We also discuss the distinct behavior of magnetic excitations in the n- and p-type cuprates.

This work was supported by NSFC Projects No. 10804047, 11274276, and A Project Funded by the Priority
Academic Program Development of Jiangsu Higher Education Institutions. CD Gong
acknowledges 973 Projects No. 2011CB922101. HQ Lin acknowledges RGC Grant of HKSAR,
Project No. HKUST3/CRF/09.


\begin{thebibliography}{99}
\bibitem{EXPLAIN} Any order with characteristic wave vector $\mathbf{Q}$
would reproduce such band structure, but the SDW is most likely.
\end{thebibliography}
\end{document}